\documentclass[aps,prd,noshowpacs,preprintnumbers,nofootinbib,floatfix,twocolumn]{revtex4}

\usepackage{bm}
\usepackage{latexsym}
\usepackage{dcolumn}
\usepackage{amsfonts,amssymb,amsmath}
\usepackage{graphicx,epsfig}
\usepackage{psfrag}

\def\eq#1{{Eq.~(\ref{#1})}}

\newcommand{\Cal}[1]{\ensuremath{\mathcal{#1}}}
\newcommand{\Alt}[6]{\ensuremath{\delta^{#1 #2 ... #3}_{#4 #5
      ... #6}}} 

\newcommand{\Altsplit}[6]{\ensuremath{\delta^{#1}_{#4}
    \delta^{#2}_{#5}... \delta^{#3}_{#6}}} 
 
\newcommand{\AltD}[6]{\ensuremath{\Cal{D}\left[^{#1 #2 ... #3}
    _{#4 #5 ... #6}\right]}}
\newcommand{\AltDsh}[4]{\ensuremath{\Cal{D}\left[^{#1 ... #2}
    _{#3 ... #4}\right]}}  
\newcommand{\AltDln}[8]{\ensuremath{\Cal{D}\left[^{#1  ... #2 #3
	... #4}_{#5 ... #6 #7 ... #8}\right]}}
\newcommand{\R}[4]{\ensuremath{R^{#1 #2}_{#3 #4}}}
\newcommand{\sgn}{{\rm sgn}}
\newcommand{\sD}[1]{\sum_{m=1}^{K}{#1}}
\newcommand{\D}[1]{\ensuremath{\Delta_{(#1)}}}
\newcommand{\tpi}{\tilde \pi}

\begin{document}


\title{Thermodynamic route to Field equations in Lanczos-Lovelock Gravity}
\author{Aseem Paranjape}
\email{aseem@tifr.res.in}
\affiliation{Tata Institute of Fundamental Research, Homi Bhabha
Road, Colaba, Mumbai - 400 005, India\\}
\author{Sudipta Sarkar}
\email{sudipta@iucaa.ernet.in}
\author{T. Padmanabhan}
\email{paddy@iucaa.ernet.in}
\affiliation{IUCAA,
Post Bag 4, Ganeshkhind, Pune - 411 007, India\\}

\date{\today}


\begin{abstract}
Spacetimes with horizons show a resemblance to thermodynamic systems
and one can associate the notions of temperature and entropy with
them. In the case of Einstein-Hilbert gravity, it is possible to
interpret Einstein's equations as the thermodynamic identity $TdS = dE
+ PdV$ for a spherically symmetric spacetime and thus provide a 
thermodynamic route to understand the dynamics of gravity. We study
this approach further and show that  the field equations
for Lanczos-Lovelock action  in a spherically symmetric
spacetime can  also be expressed as $TdS = dE + PdV$ with $S$ and $E$
being given by expressions previously derived in the literature by
other approaches. The  Lanczos-Lovelock  Lagrangians are of the form
$\Cal{L}=Q_a^{\phantom{a}bcd}R^a_{\phantom{a}bcd}$ with
$\nabla_b\,Q_a^{\phantom{a}bcd}=0$. In such models, the expansion of
$Q_a^{\phantom{a}bcd}$ in terms of the derivatives of the metric
tensor determines the structure of the theory and higher order terms
can be interpreted quantum corrections to Einstein gravity. Our result
indicates a deep connection between the thermodynamics of horizons and
the allowed quantum corrections to standard Einstein gravity, and
shows that the relation $TdS = dE + PdV$ has a greater domain of
validity that Einstein's field equations.

\end{abstract}

\maketitle
\vskip 0.5 in
\noindent
\maketitle
\section{Introduction}
There is an intriguing analogy between the gravitational dynamics of
horizons  and thermodynamics, which is not yet understood at a deeper
level \cite{paddy1}. One possible way of interpreting these results is
to assume that spacetime is analogous to an elastic solid and
equations describing its dynamics are similar to those of elasticity,
(the ``Sakharov paradigm"; see e.g., ref. \cite{sakharov}). The unknown,
microscopic degrees of freedom of spacetime (which should be analogous
to the atoms in the case of solids) will play a role only when
spacetime is probed at Planck scales (which would be analogous to the
lattice spacing of a solid). The exception to this general rule arises 
when we consider horizons \cite{magglass} which have finite
temperature and block information from a family of observers. In a
manner which is not fully understood, the horizons link certain
aspects of microscopic physics with  the bulk dynamics just as
thermodynamics can provide a link between statistical mechanics and
(zero temperature) dynamics of a solid. If this picture is correct,
then one should be able to link the equations describing bulk
spacetime dynamics with horizon thermodynamics. There have been
several approaches which have attempted to do this with different
levels of success \cite{sakharov,paddy1,paddyholo}. The most explicit
example occurs \cite{paddy2} in the case of spherically symmetric
horizons in 4-D . In this case, Einstein's equations can be
interpreted as a thermodynamic relation $TdS=dE+PdV$ arising out of
virtual displacements of the horizon.

 This result was  derived in the context of Einstein-Hilbert gravity
 arising from the Lagrangian $ L_{EH}\propto R \sqrt{- g}$. But if
 gravity is a long wavelength, emergent phenomenon then the
 Einstein-Hilbert action is just  the first term in the expansion for
 the low energy effective action. It is natural to expect
 \textit{quantum corrections} to the Einstein-Hilbert action
 functional which will, of course, depend of the nature of the
 microscopic theory but will generally involve  higher derivative
 correction terms in the Einstein-Hilbert action \cite{perturb}. In
 particular, such terms also arise in the effective low energy actions
 of string theories \cite{string}. One such higher derivative term
 which has attracted a fair amount of attention is Lanczos-Lovelock gravity
 \cite{lovelock} of which the lowest order correction appears as a
 Gauss-Bonnet term in $D\,(\,>4)$ dimensions. We study the structure
 of a general Lanczos-Lovelock type Lagrangian and the resulting equation of
 motion for a static and spherically symmetric spacetime, near a
 Killing horizon. We find that the equivalence of the equation of
 motion and the thermodynamic identity $TdS=dE+PdV$ \emph{transcends
 Einstein gravity and is applicable even in this more general case.}
 This remarkable result indicates that there is a deep connection
 between the thermodynamics of gravitational horizons, and the
 structure of the quantum corrections to Einstein gravity.

 Our result requires fairly involved combinatorial arguments and
 detailed algebra. In order not to lose the physical picture, we have
 structured the paper as follows: in the next section, we will
 briefly review the case of Einstein-Hilbert gravity to set the
 stage \cite{paddy2}. In section 3 we will display an explicit
 calculation relating the equation of motion for gravity with a
 Gauss-Bonnet correction term (the simplest non-trivial example of  Lanczos-Lovelock
 gravity) with thermodynamic quantities. In section 4 we will
 generalize the result to Lanczos-Lovelock actions with all the terms
 allowed for a given number of dimensions $D$, and discuss the
 implications in section 5. 

\section{The Einstein-Hilbert case}
 Consider a static, spherically symmetric spacetime with a horizon,
 described by the metric:  
\begin{equation}
ds^2 = -f(r) c^2 dt^2 + \frac{1}{f(r)} dr^2 +r^2 d\Omega^2.
\end{equation}
 We will assume that the function $f(r)$ has a simple zero at $r=a$
 and that $f'(a)$ is finite, so that spacetime has a horizon at $r=a$
 with non-vanishing surface gravity
 $\kappa=f^\prime(a)/2$. Periodicity in Euclidean time allows us to
 associate a temperature with the horizon  as $k_BT=\hbar
 c\kappa/2\pi=\hbar c f'(a)/4\pi$ where we have introduced normal
 units. (Even for spacetimes with multi-horizons this prescription is
 locally valid for each horizon surface.) Einstein's equation for this
 metric, $rf'(r)-(1-f)=(8\pi G/c^4) Pr^2\,$ (where $P$ is the radial
 pressure), evaluated at $r=a$ gives 
\begin{equation}
\frac{c^4}{G}\left[\frac{1}{2} f'(a)a - \frac{1}{2}\right] = 4\pi P a^2 
\label{reqa}
\end{equation} 
If we now consider two solutions with two different radii $a$ and
$a+da$ for the horizon, then multiplying \eq{reqa} by $da$, and
introducing a $\hbar$ factor \textit{by hand} into an otherwise
classical equation, we can rewrite it as 
\begin{equation}
   \underbrace{\frac{{{\hbar}} cf'(a)}{4\pi}}_{\displaystyle{k_BT}}
    \ \underbrace{\frac{c^3}{G{{\hbar}}}d\left( \frac{1}{4} 4\pi a^2
    \right)}_{ 
    \displaystyle{dS}}
  \ \underbrace{-\ \frac{1}{2}\frac{c^4 da}{G}}_{
    \displaystyle{-dE}}
 = \underbrace{P d \left( \frac{4\pi}{3}  a^3 \right)  }_{
    \displaystyle{P\, dV}}
\label{EHthermo}
\end{equation}
and read off the expressions:
\begin{equation}
 S=\frac{1}{4L_P^2} (4\pi a^2) = \frac{1}{4} {\frac{A_H}{L_P^2}}; \quad
    E={\frac{c^4}{2G}} a  
    =\frac{c^4}{G}\left( \frac{A_H}{16\pi}\right)^{1/2}
\end{equation} 
where $A_H$ is the horizon area and $L_P^2=G\hbar/c^3$. Thus
Einstein's equations can be cast as a thermodynamic identity. Three
comments are relevant regarding this result, especially since these comments are
valid for our generalization discussed in the rest of the paper as well: 

(a) The combination $TdS$ is completely classical and is independent
of $\hbar$ but $T\propto \hbar$ and $S\propto 1/\hbar$. This is
analogous to the situation in classical thermodynamics when compared
to statistical mechanics. The $TdS$ in thermodynamics is independent
of Boltzmann's constant while statistical mechanics will lead to an
$S\propto k_B$  and $T\propto1/k_B$. 

(b) In spite of superficial similarity, \eq{EHthermo} is different
from the conventional first law of black hole thermodynamics, (as well
as some previous attempts to relate thermodynamics and gravity, like
e.g. the second paper in ref. \cite{sakharov}), due to the presence of
$PdV$ term. This relation  is more in tune with the membrane paradigm
\cite{membrane} for the blackholes. This is easily seen, for example,
in the case of Reissner-Nordstrom blackhole for which $P\neq0$. If a
\textit{chargeless} particle of mass $dM$ is dropped into a
Reissner-Nordstrom blackhole, then an elementary calculation shows
that the energy defined above as $E\equiv a/2$ changes by $dE= (da/2)
=(1/2)[a/(a-M)]dM\neq dM$ while it is $dE+PdV$ which is precisely
equal to $dM$ making sure $TdS=dM$. So we need the $PdV$ term to get
$TdS=dM$ when a \textit{chargeless} particle is dropped into a
Reissner-Nordstrom blackhole. More generally, if $da$ arises due to
changes $dM$ and $dQ$, it is easy to show  that \eq{EHthermo} gives
$TdS=dM -(Q/a)dQ$ where the second term arises from the electrostatic
contribution from the horizon surface charge as expected in the
membrane paradigm. 

(c)
In standard thermodynamics, we consider two equilibrium states of a
system differing infinitesimally in the extensive variables like
entropy, energy, and volume by $dS$, $dE$ and $dV$ while having same
values for the intensive variables like temperature $(T)$ and pressure
$(P)$. Then, the first law of thermodynamics asserts that $TdS = PdV +
dE$ for these states. In a similar way, Eq.~(\ref{EHthermo}) can be
interpreted as a connection between two quasi-static equilibrium
states where both of them are spherically symmetric solutions of
Einstein equations with the radius of horizon differing by $da$ while
having same source $T_{ij}$ and temperature $T = \kappa / 2 \pi $. This
formalism does not depend upon what causes the change of the horizon
radius and is therefore very generally applicable. Note
that the structure of \eq{EHthermo} itself allows us to ``read off" the
expressions for entropy and energy. The validity of this approach as
well as the uniqueness of the resulting expressions for $S$ and $E$
are discussed at length in ref. \cite{paddy2} and will not be repeated
here.   

\section{A First correction: Gauss-Bonnet gravity}
We shall now turn our attention to the more general case. (Hereafter,
we shall adopt natural units, in which $\hbar = c = G = 1$.) A
natural generalization of the Einstein-Hilbert Lagrangian is provided
by the Lanczos-Lovelock Lagrangian, which is the sum of dimensionally
extended Euler densities, 
\begin{eqnarray}
\Cal{L}^{(D)} = \sD{c_m\Cal{L}^{(D)}_m}
\label{lovelock}
\end{eqnarray} 
where the $c_m$ are arbitrary constants and $\Cal{L}^{(D)}_m$ is the
$m^{th}$ order Lanczos-Lovelock term given by,
\begin{eqnarray}
\Cal{L}^{(D)}_m = \frac{1}{16\pi} 2^{-m} \delta^{a_1 b_1 ...a_m
  b_m}_{c_1 d_1 ...c_m d_m}  R^{c_1 d_1}_{a_1 b_1}...R^{c_m
  d_m}_{a_m b_m},
\label{LLII}
\end{eqnarray}
where $R^{ab}_{cd}$ is the $D$-dimensional Riemann tensor, and
the generalized alternating (`determinant') tensor
$\delta^{\cdots}_{\cdots}$  is totally antisymmetric in both set of 
indices. For $D=2m$, $16\pi\Cal{L}^{(2m)}_m$ is the Euler density of
the $2m$-dimensional manifold. We set $\Cal{L}_0 = 1$ and hence,
$c_0$ is proportional to the cosmological constant. The
Einstein-Hilbert Lagrangian is a special case of Eq.~(\ref{lovelock})
when only $c_1$ is non zero. These Lagrangians are free from ghosts
and are quasi-linear in nature (see the first reference in
\cite{string}). 

In this section we will concentrate on the first correction term,
namely $\mathcal{L}_2 $ which is the Gauss-Bonnet Lagrangian. In four
dimensions, this term is a total derivative while higher order
interactions are simply zero; we will work with spacetimes for which
$D > 4$. Then the relevant action functional of the theory is given
by, 
\begin{eqnarray}
\Cal{A} = \int d^{D}x \sqrt{-g} \left[\frac{1}{16\pi} ( R  + \alpha
  \mathcal{L}_{GB}) \right] + \Cal{A}_{matter} . \label{action} 
\end{eqnarray}
where $R$ is the $D$-dimensional Ricci scalar; and $\Cal{L}_{GB}$
is the Gauss-Bonnet Lagrangian which has the form, 
\begin{eqnarray}
\mathcal{L}_{GB} = R^2 - 4 R_{ab} R^{ab} + R_{abcd} R^{abcd}. 
\end{eqnarray} 
This type of action can be derived from superstring theory in the low
energy limit. In that case $\alpha$ is regarded as the inverse string
tension and is positive definite. At least in this context, it
makes sense to think of the second term as a correction to Einstein
gravity. (We have not added a cosmological constant to the action for
simplicity; all our results below trivially generalize in the
presence of a bulk cosmological constant.)  The equation of motion for
this semiclassical action in \eq{action} is given by, 
\begin{eqnarray}
G_{ab} + \alpha H_{ab} = 8 \pi T_{ab}
\end{eqnarray}
where
\begin{eqnarray}
{G}_{ab}&\equiv&R_{ab}-\frac{1}{2}g_{ab}R,\\
{H}_{ab}&\equiv&2\Bigl[RR_{ab}-2R_{aj}R^j_{~b}
-2R^{ij}R_{aibj}
\nonumber
\\
&& ~~~~
 +R_{a}^{~ijk}R_{bijk}\Bigr]
-\frac{1}{2}g_{ab}\Cal{L}_{GB}.
\end{eqnarray}
 Consider again a static spherically symmetric solution of the form
 \cite{GBsol}, 
\begin{eqnarray}
ds^2 = -f(r) dt^2 + f^{-1}(r) dr^2 + r^2 d\Omega^{2}_{D-2}.
\label{static}
\end{eqnarray}
where $d\Omega^{2}_{D-2}$ is the metric of the $D-2$ dimensional space
of constant curvature $k$. Here and below, we will work with the case
of spherical geometry with $k=1$, but all the results can easily be
generalized for $k \neq 1$. Spherical symmetry allows us to write, in
the energy momentum tensor \cite{comm1},    
$
T_{t}^{t} = T_{r}^{r} \equiv \epsilon(r)/8 \pi.
$
Then the equation of motion which determines the only nontrivial
metric component $f(r)$ is given by \cite{torii}, 

\begin{eqnarray}
rf' - (D-3) (1-f)  + \frac{\bar{\alpha}}{r^2} (1 - f) \\ \nonumber
\left[2 r f' - (D-5)(1-f)\right] &=& \frac{2  \epsilon(r)}{D-2}
r^2\label{GBeq} 
\end{eqnarray}
where $\bar{\alpha} = (d - 3)(d-4) \alpha$. Note that, $D=4$ and
$\alpha = 0$ refers to Einstein-Hilbert gravity and in this limit one
can recover Eq.~(\ref{EHthermo}). The horizon is obtained from the
location of zeroes of the function $f(r)$. In general $f(r)$ may have
several zeroes but we will concentrate locally on any one of them. Let
$r = a$ be a horizon for this spacetime with $f(r = a) =0$, the
temperature associated with this horizon being $ T = \kappa/2\pi=f'(a)
/4 \pi$. As before, we evaluate \eq{GBeq} at $r=a$ to obtain: 
\begin{eqnarray}
f'(a) \left[a + \frac{2 \bar{\alpha} }{a} \right]  -(D-3) - \frac{
  \bar{\alpha} ~ (D-5)}{a^2} = \frac{2  \epsilon(a)}{D-2} a^2
\end{eqnarray}
Our aim is to introduce in this equation a factor $dV$ and see whether 
one can read off entropy $S$ and energy $E$ from an equation of the
form $TdS=dE+PdV$. Knowing the volume element in the $D$ dimensional
space, we multiply  both sides of the above equation by the factor
$(D-2) A_{D-2} a^{D-4} da /16 \pi$, where $A_{D-1}= 2 \pi^{D/2} /
\Gamma(D/2)$ is the area of a unit $(D-1)$ sphere. Identifying the
pressure $P = T_{r}^{r}$ and the relevant volume  $V = A_{D-2}
a^{D-1} /(D-1)$ we can rewrite this equation (after some
straight forward algebra) in the form, 
\begin{align}
\frac{\kappa}{2 \pi} &d\left(\frac{A_{D-2}}{4} a^{D-2}
\left[1+\left(\frac{D-2}{D-4}\right)\frac{2~ \bar{\alpha}}{a^2}
  \right]\right)\nonumber\\
& - d \left[\frac{(D-2) A_{D-2} a^{D-3}}{16 \pi}
  \left(1 + \frac{\bar{\alpha} }{a^2} \right)\right] = P dV. 
\label{tdseqn}
\end{align}
The first term in the left hand side is in the form $TdS$ and our
analysis allows us to read off  the expression of entropy $S$ for the
horizon as,  
\begin{eqnarray}
S =\frac{A_{D-2}}{4} a^{D-2}
\left[1+\left(\frac{D-2}{D-4}\right)\frac{2~ \bar{\alpha}}{a^2}
  \right]. 
\end{eqnarray}
This is \textit{precisely} the expression for the entropy in
Gauss-Bonnet gravity calculated by several authors \cite{entropygb} by
more sophisticated methods. Further we can interpret the second term
on the left hand side of \eq{tdseqn} as $dE$ where $E$ is the  energy
of the system defined as,  
\begin{eqnarray}
E =\frac{(D-2) A_{D-2} a^{D-3}}{16 \pi} \left(1 + \frac{\bar{\alpha}
}{a^2}\right)\,,
\end{eqnarray}
which also matches with the correct expression of energy $E$ for
Gauss-Bonnet gravity without cosmological constant
\cite{entropygb,comm2}. 

This shows that our result for the Einstein-Hilbert action generalizes 
to the Gauss-Bonnet case as well and precisely reproduces the
expressions for entropy and energy (obtained in the literature by other
methods). It is clear that, at least in this context, the
thermodynamic relation transcends the Einstein field equations. We
will now show that the same result holds for the general
Lanczos-Lovelock action. 
 
\section{Gravity with the Complete Lanczos-Lovelock Action}
We now turn to the more general case of Lanczos-Lovelock gravity in $D$
dimensions with a Lagrangian given by $\Cal{L}^{(D)}=\sD{ c_m
  \Cal{L}^{(D)}_m }$ where:  
\begin{equation}
\Cal{L}^{(D)}_m=\frac{1}{16\pi}2^{-m}
\Alt{a_1}{a_2}{a_{2m}}{b_1}{b_2}{b_{2m}}
\R{b_1}{b_2}{a_1}{a_2}... \R{b_{2m-1}}{b_{2m}}{a_{2m-1}}{a_{2m}}\,,
\label{LL1}
\end{equation}
We assume that $D\geq 2K+1$ and ignore the cosmological
constant for simplicity. These Lagrangians have the peculiar property \cite{tpconf}
that their variation leads to equations of motion that are equivalent
to the ordinary \emph{partial} derivatives of the Lagrangian density
with respect to the metric components $g^{ab}$,
\begin{align}
E^a_{b} &\equiv \sD{c_m E^{a}_{b
    (m)}}=\frac{1}{2}T^a_{b}\,,\nonumber\\
E^a_{b (m)}&\equiv \frac{1}{\sqrt{-g}}g^{ai}\frac{\partial}{\partial
  g^{ib}}\left(\Cal{L}^{(D)}_m\sqrt{-g}\right)\,.
\label{LL3}
\end{align}
The factor $1/2$ with $T^a_{b}$ appears since we have normalized
$\Cal{L}^{(D)}_m$ to contain a factor of $1/(16\pi)$.

We are interested in the near-horizon structure of the
$E^t_{t}$ equation, for a spherically symmetric metric of the form
\eq{static}, and will demonstrate that this structure can (also!) be
represented as the thermodynamic identity $TdS=dE+PdV$. To this end,
we will consider the Rindler limit (see the first ref. in
\cite{paddy1}) of such a metric, by which we mean that we will study
the metric \eqref{static} near the horizon at $r=a$ and bring it to
the Rindler form  
\begin{equation}
ds^2=-N^2dt^2+\frac{dN^2}{\kappa^2+\Cal{O}(N)}+\sigma_{AB}dy^Ady^B\,,
\label{rindler}
\end{equation}
This form essentially arises by using a coordinate system in which the
level surfaces of the metric component $g_{00}$ (which vanishes on the
horizon) define the spatial coordinate $N$. The constant $\kappa$
appearing in the $g_{11}$ term above can be shown to coincide with the
surface gravity of the horizon. In this section, capitalized Latin
indices correspond to the transverse coordinates on the
$t=\,$constant, $N =\,$ constant surfaces of dimension $D-2$ and
$\sigma_{AB}$ is the metric on these surfaces. Denoting
the extrinsic curvature of these $(D-2)$-surfaces by $K_{AB}$, it is
easy to show that for the metric \eqref{static}, the Rindler limit
gives, 
\begin{subequations}
\begin{align}
\sigma_{AB}&=\sigma_{(1)AB}+\frac{N^2}{\kappa a}\sigma_{(1)AB} +
\Cal{O}(N^4) \,,
\label{rSa}\\
\sigma^{AB}&=\sigma_{(1)}^{AB}-\frac{N^2}{\kappa a}\sigma_{(1)}^{AB} + 
\Cal{O}(N^4)\,,
\label{rSb}\\
K_{AB}&=\,-\frac{N}{a}\sigma_{(1)AB}+\Cal{O}(N^2)\,,
\label{rSc}
\end{align}
\label{rindSeries}
\end{subequations}
where $\sigma_{(1)AB}=a^2\tilde \sigma_{(1)AB}$,  $\tilde
\sigma_{(1)AB}$ being the metric on a unit $(D-2)$-sphere, and
$\sigma_{(1)}^{AC}\sigma_{(1)CB}=\delta^A_{B}$.

Next we display the (near-horizon) structure of the $D$-dimensional
Riemann tensor. We will drop the superscript $D$ when considering
$D$-dimensional quantities, but retain the superscript for
$(D-2)$-dimensional quantities. It will turn out, for reasons that will
become apparent shortly, that the Riemann tensor components of the
form $R^{ti}_{jk}$ and $R^{jk}_{ti}$ will not contribute to
the $E^t_{\,t}$ equation of motion. The remaining components of
$R^{ij}_{kl}$ are,   
\begin{subequations}
\begin{align}
&R^{NA}_{NB}=\kappa\partial_NK^A_{B}+\Cal{O}(N^2)
  =\,-\frac{\kappa}{a}\delta^A_B + \Cal{O}(N^2)\,, 
\label{rMa}\\
&R^{NA}_{BC}=K^A_{C:B}-K^A_{B:C}=\Cal{O}(N)\,,
\label{rMb}\\
&R^{BC}_{NA}=g_{NN}(K_{A}^{\,\,C:B}-K_{A}^{\,\,B:C})=\Cal{O}(N)\,,
\label{rMc}\\
&R^{AB}_{CD}=\,^{(D-2)}R^{AB}_{CD} + \Cal{O}(N^2)\,,
\label{rMd}
\end{align}
\label{riemann}
\end{subequations}
where the colon denotes a covariant derivative using the
$(D-2)$ dimensional metric $\sigma_{AB}$. Also, since the $(D-2)$
dimensional hypersurfaces are maximally symmetric, their Riemann
tensor $^{(D-2)}R^{AB}_{CD}$ takes on the particularly simple form 
\begin{equation}
\,^{(D-2)}R^{AB}_{CD}=\frac{1}{a^2}\left(\delta^A_{\,C}\delta^B_{\,D}
- \delta^B_{\,C}\delta^A_{\,D}\right)\,.
\label{sphere1}
\end{equation}
With these results, we can begin analyzing the near-horizon structure
of the $E^t_{t}$ equation in Lanczos-Lovelock gravity. Since the
equation depends linearly on the terms $E^t_{t (m)}$, it is sufficient
to analyze these terms individually.  
\begin{equation}
E^t_{t (m)}= \frac{1}{\sqrt{-g}}g^{tt}\frac{\partial}{\partial
  g^{tt}}\left(\Cal{L}^{(D)}_m\sqrt{-g}\right)\,.
\label{LL4}
\end{equation}
On writing $R^{ij}_{kl}=g^{ja}R^i_{\phantom{i}akl}$, the derivative with
respect to $g^{tt}$ can be performed. Using the symmetries of the
alternating tensor, together with the relation
$(\partial\sqrt{-g})/(\partial g^{tt})=-(1/2)\sqrt{-g}g_{tt}$ and the
fact that $g_{0N}=g_{0A}=0$ for the static Rindler metric, we find,
\begin{equation}
E^t_{t(m)}=\frac{1}{16\pi}\frac{m}{2^m}
  \Alt{a_1}{a_2}{a_{2m}}{t}{b_2}{b_{2m}}
  \R{t}{b_2}{a_1}{a_2}...~\R{b_{2m-1}}{b_{2m}}{a_{2m-1}}{a_{2m}}
  -\frac{1}{2}\Cal{L}^{(D)}_m \,.
\label{LL5}
\end{equation}
We will now show that the summations involved in the first term of
\eq{LL5} are cancelled by terms in $\Cal{L}^{(D)}_m$. Let us
categorize the terms that appear in $\Cal{L}^{(D)}_m$ into those
in which the index value $t$ appears at least once, which we denote by
$\{T\}$, and those in which $t$ does not appear, which we denote by
$\{\bar T\}$. Symbolically then, $\Cal{L}^{(D)}_m=\{T\}+\{\bar T\}$. 
In the case of standard Einstein gravity, we have
$16\pi\Cal{L}^{(D)}_1=R=2R^t_{t}+R^{\alpha\beta}_{\alpha\beta}$ 
and one clearly recognizes $2R^t_{t}$ as $\{T\}$. Since the
Einstein tensor is $G^t_{\,t}=R^t_{\,t}-(1/2)R$, the terms $\{T\}$ in
$R$ are precisely cancelled in $G^t_{\,t}$. We will now show that
exactly the same feature occurs in the $m^{th}$ Lanczos-Lovelock
case. To see this, we construct the set $\{T\}$ as follows. Focussing
on the lower row of the alternating tensor in the expression
\eqref{LL1} for $\Cal{L}^{(D)}_m$, we have $2m$ choices for the
location of the index value $t$. Due to the symmetries of the
alternating tensor and the Riemann tensor, each choice results in the
same term, and we can write, 
\begin{equation}
\{T\}=\frac{2}{16\pi}\frac{m}{2^m}\Alt{a_1}{a_2}{a_{2m}}{t}{b_2}{b_{2m}}
\R{t}{b_2}{a_1}{a_2}...~\R{b_{2m-1}}{b_{2m}}{a_{2m-1}}{a_{2m}}\,.
\label{LL6}
\end{equation} 
A comparison with \eq{LL5} shows that the first term in that equation
is simply $(1/2)\{T\}$, and we are left with
\begin{equation}
E^t_{t(m)}=\,-\frac{1}{2}\{\bar T\}\,.
\label{LL7}
\end{equation}
Note that the set $\{\bar T\}$ is not \emph{a priori} a null set since
we have assumed $D\geq 2m+1$.  To simplify the contribution of $\{\bar
T\}$, we further split this set as follows. This set contains terms
with exactly one occurence of the index value $N$, denoted $\{\bar T,
1\,N\}$, terms with two occurences of $N$, denoted $\{\bar T, 2\,N\}$,
and terms with no occurences of $N$, denoted $\{\bar T, \bar
N\}$. (The total antisymmetry of the alternating tensor forbids more
than one occurence of $N$ in any row.) Each term in the set $\{\bar T,
1\,N\}$ contains one factor of the type $R_{NA}^{BC}$ or
$R_{BC}^{NA}$, and \eq{riemann} shows that these terms are
$\Cal{O}(N)$ and don't contribute on the horizon. Similarly, the set
$\{\bar T, 2\,N\}$ contains one type of terms in which the two $N$'s
appear in \emph{different} factors of
$R^{\alpha\beta}_{\mu\nu}$. These terms contain two factors each
of $R_{NA}^{BC}$ or $R_{BC}^{NA}$, rendering these terms
$\Cal{O}(N^2)$. The contribution from $\{\bar T, 2\,N\}$ reduces to
the $4m$ identical terms in which both the $N$'s appear in the
\emph{same} factor of $R^{\alpha\beta}_{\mu\nu}$, and is given by, 
\begin{align}
&\{\bar T, 2\,N\}=\nonumber\\
&\frac{4}{16\pi}\frac{m}{2^m}\Alt{N}{A_2}{A_{2m}}{N}{B_2}{B_{2m}}    
\R{N}{B_2}{N}{A_2}...~\R{B_{2m-1}}{B_{2m}}{A_{2m-1}}{A_{2m}}
+\Cal{O}(N^2)\,, 
\label{LL8}
\end{align}
where $A_2,B_2,...=y^A$. The set $\{\bar T, \bar N\}$ will not be
\emph{a priori} a null set whenever $D\geq2m+2$, and its contribution
is, 
\begin{align}
\{\bar T, \bar N\}&=
\frac{1}{16\pi}\frac{1}{2^m}\Alt{A_1}{A_2}{A_{2m}}{B_1}{B_2}{B_{2m}}
\R{B_1}{B_2}{A_1}{A_2}...~\R{B_{2m-1}}{B_{2m}}{A_{2m-1}}{A_{2m}}
\nonumber\\
&= \Cal{L}^{(D-2)}_m + \Cal{O}(N^2)\,,
\label{LL9}
\end{align}
where we have used \eq{rMd} and recognized the structure of
$\Cal{L}^{(D-2)}_m$ in the resulting term. Finally, using
Eqs. \eqref{LL8} and \eqref{LL9}, substituting for the near-horizon
structure of $\R{N}{B}{N}{A}$ from \eq{riemann} and relabelling some
indices, we find, 
\begin{align}
&E^t_{t(m)}=\nonumber\\
&\frac{\kappa  m}{16\pi}\frac{1}{2^{m-1}}
\left(\frac{1}{a}\delta_{\,A_1}^{B_1}\right)
\Alt{N}{A_1}{A_{2m-1}}{N}{B_1}{B_{2m-1}}
\,^{(D-2)}R^{B_2B_3}_{A_2A_3}...\nonumber\\ 
&~~~~~~~~~~~~~~~~~~~~~~~~~-\frac{1}{2}\Cal{L}^{(D-2)}_m +
\Cal{O}(N)\,,\label{LL10}
\end{align} 
where $\Cal{L}^{(D-2)}_m$ will contribute only when $D\geq 2m+2$. The
first term of \eq{LL10} can be simplified by noting the following. The
alternating tensor $\Alt{N}{A_1}{A_{2m-1}}{N}{B_1}{B_{2m-1}}$ can be
replaced by $\Alt{A_1}{A_2}{A_{2m-1}}{B_1}{B_2}{B_{2m-1}}$ since
$\delta^N_{\,N}=1$ and $\delta^N_{\,A}=0$. Further, due to the total
antisymmetry of the alternating tensor, each factor of
$^{(D-2)}R^{AB}_{CD}$ can be replaced by
$(2/a^2)\delta^A_{\,C}\delta^B_{\,D}$, and there are $(m-1)$ such
factors. Putting everything together, we find,  
\begin{align}
&E^t_{t(m)}=\nonumber\\
&\frac{\kappa m}{16\pi}\frac{1}{a^{2m-1}}
 \left( \Alt{A_1}{A_2}{A_{2m-1}}{B_1}{B_2}{B_{2m-1}}\right)
  \Altsplit{B_1}{B_2}{B_{2m-1}}{A_1}{A_2}{A_{2m-1}} \nonumber\\
&~~~~~~~~~~~~~~~~~~~~~~~~~-\frac{1}{2}\Cal{L}^{(D-2)}_m +
\Cal{O}(N)\,.
\label{sphere4}
\end{align}
We can further perform the summations over $A_1$ and $B_1$ and
rearrange terms to obtain,
\begin{align}
\frac{\kappa m}{8\pi}\frac{D-2m}{a^{2m-1}}
 &\left( \Alt{A_2}{A_3}{A_{2m-1}}{B_2}{B_3}{B_{2m-1}}\right)
  \Altsplit{B_2}{B_3}{B_{2m-1}}{A_2}{A_3}{A_{2m-1}} \nonumber\\
&=2E^t_{t(m)}+\Cal{L}^{(D-2)}_m + \Cal{O}(N)\,.
\label{sphere5}
\end{align}
We have relegated the proof of \eq{sphere5} to the appendix, since it
involves combinatorial arguments and is rather involved. 

We are now ready to make the connection with the thermodynamic
identity by a procedure which is \textit{identical} to that used in the
Einstein-Hilbert and Gauss-Bonnet cases. We wish to multiply the
$E^t_{\,t}$ equation of motion $E^t_{\,t}=(1/2)T^t_{\,t}$ evaluated
\emph{on the horizon} by the volume differential $dV=A_{D-2}a^{D-2}da$
and try to ``read off'' expressions for the entropy, energy, etc. We
note that multiplying \eq{sphere5} by the coupling constant $c_m$ and
summing over $m$ will give $2E^t_{\,t}$ as the first term on the right
hand side, which we can replace by $T^t_{\,t}$. We also know that in
the spherically symmetric case we have
$T^t_{t}=T^r_{r}=T^N_{N}=P$ with $P$ the radial
pressure. The equation obtained after these replacements will be
equivalent to the equation of motion and multiplying it with $dV$ will
result in the following, 
\begin{align}
\frac{\kappa}{2\pi}d&\left(\sD{\frac{m}{4}c_mA_{D-2}a^{D-2m}
  \left(\Alt{A_2}{.}{A_{2m-1}}{B_2}{.}{B_{2m-1}}\right) 
\delta^{B_2}_{A_2}...\delta^{B_{2m-1}}_{A_{2m-1}} }\right) \nonumber\\  
&=PdV+\sD{c_mA_{D-2}a^{D-2}\Cal{L}^{(D-2)}_m da}\,.
\label{sphere6}
\end{align}
Recognizing $\kappa/2\pi$ as the temperature $T$, we are forced to
identify the quantity inside parentheses on the left hand side above,
as the entropy $S$. Noting that the alternating tensor that appears
here contains $2m-2$ indices per row, and recalling the simple
structure of the $(D-2)$ dimensional Riemann tensor from \eq{sphere1},
we can rewrite our entropy as $S=\sD{S^{(m)}}$ with $S^{(m)}$ given by,
\begin{align}
S^{(m)}&=4\pi mc_mA_{D-2}a^{D-2}\Cal{L}^{(D-2)}_{m-1}\nonumber\\
&=4\pi mc_m\int_\Cal{H}{\Cal{L}^{(D-2)}_{m-1}\sqrt\sigma
  d^{D-2}y}\,,
\label{LLent1}
\end{align}
Note that, in our approach, which is identical to what we followed in
the case of Einstein-Hilbert and Gauss-Bonnet cases, we have no choice
in the expression for $S$. Remarkably enough, this is \emph{precisely}
the entropy of the horizon in Lanczos-Lovelock gravity which has been
computed by several authors (see, e.g., the first ref. in
\cite{entropygb}).  

Having identified the
$TdS$ and $PdV$ terms in \eq{sphere6}, we ask whether the remaining
quantity can be interpreted as the differential of some function. We
find that this is indeed the case and we have,
\begin{align}
\sD{c_mA_{D-2}a^{D-2}\Cal{L}^{(D-2)}_m da}&=d\left(\sD{c_mE_{(m)}}
\right)\,,
\label{energy1}\\
E_{(m)}= \frac{1}{16\pi}A_{D-2}a^{D-(2m+1)}&\prod_{j=2}^{2m}{(D-j)}\,,
\label{energy2}
\end{align}
The proof of \eq{energy1} can be found in the appendix. This requires
us (again we have no choice in the matter!) to interpret the quantity 
$E=\sD{c_mE_{(m)}}$ as the energy associated with the horizon;
incredibly enough, we find that \emph{exactly} this expression has
been computed by other authors \cite{cai} as the energy of the horizon
in spherically symmetric Lanczos-Lovelock gravity! 

Incidentally, the expression for the differential of the energy $dE$
in all the cases presented here shows that this contribution arises
from the term $\Cal{L}^{(D-2)}_m$ (which, for the Einstein-Hilbert
case in $D=4$ for example, is simply $\,^{(2)}R$). Thus the energy
associated with the horizon originates in the transverse geometry of
the horizon.  

We have therefore proved that, for the spherically symmetric case, the
equation of motion $E^t_{t}=(1/2)T^t_{t}$ can be recast in the
form 
\begin{equation}
\left(\frac{\kappa}{2\pi}\right)dS~=dE+PdV \,.
\label{finale}
\end{equation}
with the differentials being interpreted as arising due to a change in
the radius of the horizon. In principle, the corrections to the
entropy and the energy coming from the higher order Lanczos-Lovelock terms
need not have preserved the structure of the first law of
thermodynamics apparent above in the gravitational field equations. 

We find it rather far-fetched to believe that this precise analogy of
the field equations with the first law of thermodynamics (albeit for
the spherically symmetric case) is a mere coincidence. This feature of
the field equations seems to point towards a deeper principle which is
yet to be understood.

\section{Discussion}
The fact that the expression for entropy $(S)$ and energy $(E)$
 obtained from this approach, by casting the equation in the form $TdS
 = dE + PdV$, \textit{matches exactly} with the standard quantum field
 theory calculations as in the case of Einstein-Hilbert gravity, is
 nontrivial and intriguing. However, it  resonates well with an
 alternative perspective on gravity which was developed in a series of
 recent papers \cite{newper,tpconf}. This alternative paradigm views
 semiclassical gravity as  based on  a generic Lagrangian of the form
 $L=Q_a^{\phantom{a}bcd}R^a_{\phantom{a}bcd}$ with
 $\nabla_b\,Q_a^{\phantom{a}bcd}=0$. The expansion of
 $Q_a^{\phantom{a}bcd}$ in terms of the derivatives of the metric
 tensor determines the structure of the theory uniquely. The zeroth
 order term gives the Einstein-Hilbert action and the first order
 correction is given by the  Gauss-Bonnet action. More importantly,
 \textit{any} such Lagrangian can be decomposed into a surface and
 bulk terms as 
$\sqrt{-g}L=\sqrt{-g}L_{bulk}+L_{sur}$ where
\begin{eqnarray}
L_{\rm bulk} &=& 2 \, Q_a^{\phantom{a}bcd} \Gamma^a_{dk}
\Gamma^k_{bc};\nonumber\\ 
L_{sur}&=&\partial_c[\sqrt{-g}V^c];\quad
V^c=2Q_a^{\phantom{a}bcd}\Gamma^a_{bd}
\end{eqnarray} 
Obviously, both $L_{sur}$ and $L_{bulk}$ contain the same information
in terms of $Q_a^{\phantom{a}bcd}$ and hence can \textit{always} be
related to each other \cite{tpconf,ayan}. It is easy to verify, for
example \cite{comm3}, that 
\begin{equation}
L=\frac{1}{2}R^a_{\phantom{a}bcd}\left(\frac{\partial V^c}{\partial
  \Gamma^a_{bd}}\right); 
\quad
L_{bulk}=\sqrt{-g}\left(\frac{\partial V^c}{\partial \Gamma^a_{bd}}\right)
\Gamma^a_{dk}\Gamma^k_{bc}
\end{equation}  
Thus the knowledge of the functional form of $L_{sur}$ or ---
equivalently --- the $V^c$ allows us to determine $L_{bulk}$ and even
$L$. (The first relation also shows that  $(\partial V^c/\partial
\Gamma^a_{bd})$ is generally covariant in spite of the appearance.)
These relations make the actions based on
$L=Q_a^{\phantom{a}bcd}R^a_{\phantom{a}bcd}$ with
$\nabla_b\,Q_a^{\phantom{a}bcd}=0$ intrinsically ``holographic" with the
surface term containing an equivalent information as the bulk. What is
more, one can show that the surface term leads to the Wald entropy in
spacetimes with horizon \cite{paddyholo,ayan}.  Since Lanczos-Lovelock
Lagrangians have this structure, it is quite understandable that the
semiclassical equations of motion have a thermodynamic interpretation. 

We can summarize the broader picture as follows:  Any geometrical
description of gravity  that obeys the principle of equivalence and is
based on  a nontrivial metric,  will allow for the propagation of
light rays to be affected by gravity. This, in turn, leads to regions
of spacetime which are causally inaccessible to classes of
observers. (These two features are reasonably independent of the
precise field equations which determine the metric.). The
inaccessiblity of regions of spacetime leads to association of entropy
with spacetime horizons. Such a point of view suggests that there will
exist a thermodynamic route to the description of gravitational
dynamics in \textit{any} metric theory which satisfies the principle
of equivalence. So the thermodynamic interpretation of gravity,
encoded in the identity $TdS = PdV + dE$, should be fairly generic and
the  semiclassical corrections to gravity --- arising from the correct
microscopic theory --- should preserve the form of this identity (with
only the expressions for $S$ and $E$ getting quantum corrections.) We
have shown that this is indeed the case for spherically symmetric
horizons in the Lanczos-Lovelock lagrangian. Such an interpretation
offers a new outlook towards the dynamics of gravity and might provide 
valuable clues regarding the nature of quantum gravity. 

\section*{Acknowledgements}
One of the authors (SS) is supported by the Council of Scientific \&
Industrial Research, India. SS also thanks Sanjay Jhingan for
discussions. AP thanks all the members of IUCAA for their warm
hospitality and pleasant company during his stay there, during which
this work was completed. 

\appendix
\section{}
\noindent
In this appendix we shall prove equations \eqref{sphere5} and
\eqref{energy1}. In order to prove \eq{sphere5}, it is sufficient to
show that,
\begin{align}
\D{k}\equiv&\left( \Alt{T_1}{T_2}{T_{k}}{B_1}{B_2}{B_{k}}\right)
  \Altsplit{B_1}{B_2}{B_k}{T_1}{T_2}{T_{k}}\nonumber\\
&=\left(D-(k+1)\right) \left(
  \Alt{T_2}{T_3}{T_{k}}{B_2}{B_3}{B_{k}}\right)
  \Altsplit{B_2}{B_3}{B_{k}}{T_2}{T_3}{T_{k}}\nonumber\\
&=\left(D-(k+1)\right)\D{k-1}\,,
\label{app1}
\end{align}
with $k=2m-1$, and the indices $T_1, B_1$, etc. ranging over the
$(D-2)$ values $2,3...(D-1)$. Having proved this, a simple
rearrangement of terms in \eq{sphere4} leads to \eq{sphere5}. We will
prove this result for a general $k$ since it will come in handy when
proving \eq{energy1}. To simplify notation we introduce the
following symbol for a single product of Kronecker delta's,
\begin{equation}
\AltD{T_1}{T_2}{T_k}{B_1}{B_2}{B_k}\equiv
\Altsplit{T_1}{T_2}{T_k}{B_1}{B_2}{B_k} \,.
\label{app2}
\end{equation}
The alternating tensor is normalized to take values $0,1$ and $-1$. In
practice, this can be done by antisymmetrizing only the upper row of
indices and we can write,
\begin{equation}
\Alt{T_1}{T_2}{T_k}{B_1}{B_2}{B_k}= \sum_{\pi\in S_{(k)}}{
  \sgn(\pi)
  \AltD{\pi(T_1)}{\pi(T_2)}{\pi(T_{k})}{~~B_1}{~~B_2~~}{~~B_{k}}
  }  \,,
\label{app3}
\end{equation}
where $S_{(k)}$ is the set of permutations of $k$ objects and 
$\sgn(\pi)$ denotes the signature of the permutation $\pi$. Our goal
is to perform the summations over the indices $T_1$ and $B_1$ in the
quantity \D{k} defined in \eqref{app1}. To simplify this
computation, we can split up the set $S_{(k)}$ as the union of sets
$S_{(k)}^j$ with $1\leq j\leq k$ where $S_{(k)}^j$ is the set of
permutations $\pi$ which map $T_j$ to $T_1$, i.e.,
\begin{equation}
S_{(k)}^j=\left\{\pi\in S_{(k)}~\mid~\pi(T_j)=T_1\right\}\,.
\label{app4}
\end{equation}
Noting that for $\pi\in S_{(k)}^1$, $\pi(T_1)=T_1$, we can write,
\begin{align}
\D{k}&=\delta^{T_1}_{B_1}\left(\Alt{T_2}{.}{T_k}{B_2}{.}{B_k}\right)
\AltD{B_1}{B_2}{B_k}{T_1}{T_2}{T_k} \nonumber\\
&+\sum_{j=2}^{k}\sum_{\pi\in S_{(k)}^j}{\sgn(\pi)
  \AltD{\pi(T_1)}{\pi(T_2)}{\pi(T_{k})}{~~B_1}{~~B_2~~}{~~B_{k}}
  \AltD{B_1}{B_2}{B_k}{\,T_1}{\,T_2}{\,T_k}}\nonumber\\
&=\left(D-2\right)\D{k-1} + \sum_{j=2}^{k}{\Cal{M}^j_{(k)}}\,,
\label{app5}
\end{align}
where the last line defines the quantities $\Cal{M}^j_{(k)}$ for $2\leq
j\leq k$. For a particular value of $j$ we get,
\begin{align}
&\Cal{M}^j_{(k)}=\nonumber\\
&\sum_{\pi\in S_{(k)}^j}{\sgn(\pi)
  \AltDln{\pi(T_1)}{T_1}{\pi(T_{j+1})}{\pi(T_k)}
	 {~~B_1~}{B_j}{~B_{j+1}~}{~B_k}}
 \AltD{B_1}{B_2}{B_k}{\,T_1}{\,T_2}{\,T_k}\,,
\label{app6}
\end{align}
where we have set $\pi(T_j)=T_1$. We now have a simple product of
Kronecker delta's for each $\pi\in S_{(k)}^j$ with $T_1$ and $B_1$
appearing explicitly. Performing the summations over $T_1$ and $B_1$
reduces this to,
\begin{align}
&\Cal{M}^j_{(k)}=\nonumber\\
&\sum_{\pi\in S_{(k)}^j}{\sgn(\pi)
  \AltDln{\pi(T_1)}{\pi(T_{j-1})}{\pi(T_{j+1})}{\pi(T_k)}
       {~~B_j~}{\,B_{j-1}}{~~B_{j+1}~~~}{~~B_k}}
 \AltDsh{B_2}{B_k}{\,T_2}{\,T_k}\,.
\label{app7}
\end{align}
From the definition of $S_{(k)}^j$, for each $\pi\in S_{(k)}^j$, the
ordered set
$\Cal{P}_\pi=\{\pi(T_1),\pi(T_2)...,\pi(T_{j-1}),\pi(T_{j+1}),...,\pi(T_k)\}$
is  simply a rearrangement of the ordered set
$\Cal{P}=\{T_2,T_3,...,T_j,...,T_k\}$. Hence there exists a one-to-one
mapping between $S_{(k)}^j$ and the set $S_{(k-1)}$ of permutations of
$(k-1)$ objects. We would like to replace the summation $\sum_{\pi\in
  S_{(k)}^j}$ by the summation  $\sum_{\tpi\in S_{(k-1)}}$. To ensure
that each term in the summation retains its correct signature after
this replacement, we must introduce an overall factor of $\sgn(C_j)$,
which is the signature of the permutation $C_j\in S_{(k)}^j$ that is
mapped to the identity of $S_{(k-1)}$. It is easy to see that this
permutation is the semicyclic rearrangement given by
$\{T_1,T_2,...,T_j,...,T_k\}\to \{T_2,T_3,...,T_j,T_1,T_{j+1}...,T_k\}$,
which has signature $(-1)^{j+1}$. We can write,
\begin{align}
&\Cal{M}^j_{(k)}=(-1)^{j+1}\times\nonumber\\
&\sum_{\tpi\in S_{(k-1)}}{\sgn(\tpi)
  \AltDln{\tpi(T_2)}{\tpi(T_{j})}{\tpi(T_{j+1})}{\tpi(T_k)}
       {~~B_j~}{\,B_{j-1}}{~B_{j+1}~~}{~~B_k}}
 \AltDsh{B_2}{B_k}{\,T_2}{\,T_k}\,.
\label{app8}
\end{align}
The order of the first $j$ indices in the lower row of the first
factor of $\Cal{D}$ in \eqref{app8} is not in the standard form. To
get $B_2$ below $\tpi(T_2)$ and so on, we simply perform the cyclic
permutation $\{B_j,B_2,...,B_{j-1}\}\to\{B_2,B_3,...B_{j-1},B_j\}$,
with the other indices left untouched. This can be done since
permutations of the upper indices in the alternating tensor are
equivalent to those of the lower indices \cite{comm4}. The permutation
introduces a factor of $(-1)^j$, which combines with the $(-1)^{j+1}$
in \eqref{app8} to give an overall factor of $(-1)$. We now find that, 
\begin{align}
\Cal{M}^j_{(k)}&=\,-\sum_{\tpi\in S_{(k-1)}}{\sgn(\tpi)
  \AltDsh{\tpi(T_2)}{\tpi(T_k)}
       {~~B_2~}{~~B_k}}
 \AltDsh{B_2}{B_k}{\,T_2}{\,T_k}\nonumber\\
&=\,-\D{k-1}
\label{app9}
\end{align}
independent of $j$. Since there are $(k-1)$ such terms, \eqref{app5}
gives us the required result, namely,
\begin{equation}
\D{k}=\left(D-(k+1)\right)\D{k-1}\,.
\label{app10}
\end{equation}
Setting $k=2m-1$ completes the proof of \eq{sphere5}. The result in
\eqref{app10} also allows us to prove \eq{energy1} in the following
way. Using arguments similar to those presented below \eq{LL10}
and evaluating all quantities on the horizon, the left hand side of
\eq{energy1} for a single value of $m$ can be expanded to give,
\begin{align}
&c_m A_{D-2} a^{D-2}\Cal{L}^{(D-2)}_m\nonumber\\
&=\frac{c_mA_{D-2}}{16\pi}\frac{a^{D-2}}{2^m}
\left(\Alt{A_1}{A_2}{A_{2m}}{B_1}{B_2}{B_{2m}}\right)
\,^{(D-2)}R^{B_1B_2}_{A_1A_2}~... \nonumber\\
&=\frac{c_m}{16\pi}A_{D-2}a^{D-(2m+2)}
\left(\Alt{A_1}{A_2}{A_{2m}}{B_1}{B_2}{B_{2m}}\right)
\AltD{B_1}{B_2}{B_{2m}}{A_1}{A_2}{A_{2m}}\nonumber\\
&=\frac{c_m}{16\pi}A_{D-2}a^{D-(2m+2)}\D{2m}\nonumber\\
&=\frac{c_m}{16\pi}A_{D-2}a^{D-(2m+2)} (D-(2m+1))\D{2m-1} \nonumber\\
&=\frac{c_m}{16\pi}A_{D-2}a^{D-(2m+2)} (D-(2m+1))
\prod_{j=2}^{2m}{(D-j)} \nonumber\\
&=\frac{dE_{(m)}}{da}\,.
\label{app11}
\end{align}
where we have recursively used \eqref{app10} to obtain the last but
one equality, and used \eq{energy2} in writing the last
equality. This completes the proof of \eq{energy1}, and consequently
of the result \eq{finale}.


\begin{thebibliography}{20}
\bibitem{paddy1}
For a recent review, see:
T. Padmanabhan, Phys.Rept. {\bf 406} , 49-125, (2005), [gr-qc/0311036];
\textit{Mod.Phys.Letts.} \textbf{A  17}, 923 (2002) [gr-qc/0202078].

\bibitem{sakharov}
A. D. Sakharov,  Sov. Phys. Dokl. {\bf 12}, 1040 (1968);
T.~Jacobson,  Phys. Rev. Lett. \textbf{75},  1260 (1995);
T.~Padmanabhan, Mod. Phys. Lett. \textbf{A 17}, 1147 (2002) [hep-th/0205278];
                \textbf{18}, 2903 (2003) [hep-th/0302068];
              Class.Quan.Grav. \textbf{21}, 4485 (2004) [gr-qc/0308070];
G.E. Volovik, Phys.Rept., \textbf{351}, 195 (2001);
G.~E. Volovik, \textit{The universe in a helium droplet}, (Oxford
University Press, 2003);  
B.L. Hu, gr-qc/0503067 and references therein.

\bibitem{magglass}
 See e.g.,T.Padmanabhan, Phys. Rev. Letts. ,  81 , 4297 (1998)
 [hep-th-9801015]; 
                   Phys. Rev. D., 59, 124012 (1999) [hep-th-9801138]
 and references therein. 



\bibitem{paddyholo}
  T. Padmanabhan, Brazilian Jour.Phys. (Special Issue)  35, 362 (2005)
  [gr-qc/0412068]; 
Mod.Phys.Letts. A , {\bf 17}, p. 1147 (2002) [hep-th/0205278]; 
                                   \textbf{18}, 2903 (2003) [hep-th/0302068];
 Gen.Rel.Grav., 34 2029-2035 (2002) [gr-qc/0205090];
Gen.Rel.Grav., 35, 2097-2103 (2003).


\bibitem{paddy2}
T. Padmanabhan , Class.Quan.Grav. {\bf 19}, 5387 (2002). [gr-qc/0204019].

\bibitem{perturb}
G. 't Hooft and M Veltman. Ann. Inst. Henri Poincare {\bf 20}, 69,
(1979); S. Deser and P. van Nieuwenhuizen, Phys. Rev. D {\bf 10}, 401,
(1974); M. H. Goroff and A Sagnotti, Nucl. Phys {\bf B 266}, 709,
(1986). 

\bibitem{string}
See e.g., B. Zwiebach, Phys. Letts. \textbf{B}, 156, 315 (1985);
M.B.~Green, J.H.~Schwarz and E.~Witten, {\it Superstring
  Theory}. (Cambridge University Press, 1987). 

\bibitem{lovelock}
C. Lanczos, Z.Phys. \textbf{73}, 147 (1932); Annals Math. \textbf{39}, 842 (1938);
D. Lovelock, Jour. Math. Phys., \textbf{12}, 498 (1971). For a recent
review, see Nathalie Deruelle and John Madore, [gr-qc/0305004], (2003). 


\bibitem{membrane}
Kip Thorne et.al (Eds.), \textit{Black Holes : The Membrane Paradigm}, Yale
University Press (1986). 


\bibitem{hawking}
N.~D. Birrel, P.~C.~W. Davies, \textit{Quantum Field Theory in Curved
Space-Time,} Cambridge University Press, Cambridge, 1982;  G.~Gibbons,
S.~Hawking, Phys. Rev. D {\bf 15}, 2738--2751, (1977); J.~M. Bardeen,
B.~Carter, and S.~W. Hawking, Commun. Math. Phys., {\bf 31}, 161--170,
(1973); R.~M. Wald,   Living
Rev. Rel.  {\bf 4}, 6, (2001). 




\bibitem{GBsol}
Wheeler J T, Nucl, Phys. {\bf B}, 268, 737; D. G. Boulware and
S. Deser, Phys. Rev. Lett. {\bf 55}, 2646, (1985). 

\bibitem{comm1}
Medved \emph{et al} \cite{DBH} have shown that this result holds on
the Killing horizon even for a general static spacetime. 



\bibitem{torii}
Takashi Torii and Hideki Maeda, Phys. Rev. D {\bf 72}, 401, (2005).

\bibitem{entropygb}
There is an extensive literature on this topic. For a sample, see
Jacobson T and Myers R C , Phys. Rev. Lett. {\bf 70}, 3684, (1993);
Myers R C and Simon J Z , Phys. Rev. D, {\bf 38}, 2434, (1998);
Rong-Gen Cai, Phys. Rev. D {\bf 65}, 084014, (2002),
Phys. Rev. Lett. {\bf 582}, 237, (2004);  S Nojiri, S D Odintsov
and S. Ogushi, Phys. Rev. D {\bf 65}, 023521, (2002) [hep-th/0108172];
S Nojiri and S D Odintsov, Phys. Lett. B521, 87, (2001) [hep-th/
  0109122]; M Cvetic, S Nojiri and S D Odintsov, Nucl. Phys. B628,
295, (2002) [hep-th/0112045]; Tim Clunan, Simon F Ross and Douglas J
Smith, Class. Quantum Grav. {\bf 21}, 3447-3458, (2004); I P Neupane,
Phys. Rev. D {\bf 67}, 061501, (2003); Y M Cho and I P Neupane,
Phys. Rev. D {\bf 66}, 024044, (2002) [hep-th/0202140]; Nathalie
Deruelle, Joseph Katz and Sachiko Ogushi, Class. Quant. Grav. {\bf
  21}, 1971, (2004) [gr-qc/0310098]; Georgios Kofinas and Rodrigo
Olea, [hep-th/0606253], (2006). For related work see Rong-Gen Cai and
Sang Pyo Kim, JHEP 0502:050, (2005) [hep-th/0501055]; M Akbar and
Rong-Gen Cai, Phys. Lett. B635, 7, (2006) [hep-th/0602156]. 

\bibitem{comm2}
Incidentally, the entire analysis can easily generalized for a
non-zero $\Lambda$; in  that case the expression of energy, for
example, will pick up an additional term $(a^2/l^2)$ where $\Lambda =
- (D-1)(D-2)/2 l^2$ etc. The cosmological constant can be thought of
as the ``zeroth'' term in the Lanczos-Lovelock series, and hence this
is a special case of the more general result for energy to be shown in
the section 4. 

\bibitem{tpconf}
 T. Padmanabhan, \textit{Dark Energy: Mystery of the Millennium}; section 5 
 Lecture  at Albert Einstein Century International Conference, Paris,
 France, 18-23 July, 2005 
  [astro-ph/0603114];
  T. Padmanabhan, \textit{Gravity: A New Holographic Perspective } 
  Lecture at the International Conference on Einstein's Legacy in the
 New Millennium, December 15 - 22, 2005, Puri, India, 
   [gr-qc/0606061].  


\bibitem{newper}
T. Padmanabhan, Int.J.Mod.Phys., D14,2263-2270 (2005) [gr-qc/0510015].
  
 

\bibitem{DBH}
A.J.M. Medved, Damien Martin and Matt Visser, Class.Quant.Grav., 21,
3111, (2004) [gr-qc/0402069].

\bibitem{cai}
Rong-Gen Cai, Phys. Lett. B582, 237, (2004) [hep-th/0311240].


\bibitem{comm3} In all these expressions, we treat
  $[g^{ab},\Gamma^a_{bc},R^a_{\ bcd}]$ as independent variables. While
  taking the derivatives, all components of $\Gamma^k_{cb}$ are
  treated as independent with no symmetry requirements. After the
  derivative is computed, one relates the $\Gamma^k_{cb}$s to the
  metric in the usual manner. Hence the order of lower indices of
  $\Gamma^k_{cb}$s in $L_{bulk}$ etc. is important.  

\bibitem{ayan}
Ayan Mukhopadhyay and T. Padmanabhan (in preparation).

\bibitem{comm4}
Alternatively, one can define a map from   $S_{(k-1)}$ to a copy of
itself such that for each   $\tpi\in S_{(k-1)}$,
$\{\tpi(T_2),\tpi(T_3) ,...,\tpi(T_j)\}\to\{\tpi(T_j),\tpi(T_2),
...,\tpi(T_{j-1})\}$. This will also pick up a factor $(-1)^j$ and
have the same effect as permuting the $B_i$'s.

\end{thebibliography}
\end{document}